**Short Paper**

# Icarus: An Android-Based Unmanned Aerial Vehicle (UAV) Search and Rescue Eye in the Sky

Manuel Luis C. Delos Santos
Asian Institute of Computer Studies, Philippines
manuelluis.delossantos@aics.edu.ph
ORCID: 0000-0002-6480-3377
(corresponding author)

Jerum B. Dasalla
Philippine State College of Aeronautics
jerumbdasalla@philsca.edu.ph
ORCID: 0000-0002-8999-9900

Jomar C. Feliciano
Asian Institute of Computer Studies, Philippines
jomarfeliciano19@gmail.com

Dustin Red B. Cabatay
Asian Institute of Computer Studies, Philippines
dustinredcabatay@gmail.com

## Abstract

*Purpose* – The purpose of this paper is to develop an unmanned aerial vehicle (UAV) using a quadcopter with the capability of video surveillance, map coordinates, a deployable parachute with a medicine kit or a food pack as a payload, a collision warning system, remotely controlled, integrated with an android application to assist in search and rescue operations.

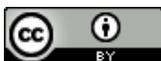

*Method* – Applied research for the development of the functional prototype, quantitative and descriptive statistics to summarize data by describing the relationship between variables in a sample or population (Kaur et al., 2018). The quadcopter underwent an evaluation using a survey instrument to test its acceptability using predefined variables to select respondents within Caloocan City and Quezon City, Philippines.

*Results* – Demographic profiles and known issues and concerns were answered by 30 respondents. The results were summarized and distributed in Tables 1 and 2.

*Discussion* - In terms of demographic profiles, the number of SAR operators within the specified areas is distributed equally, most are male, single, and within the age bracket of 31 and above. In issues and concerns, the most common type of search and rescue was ground search and rescue. Human error is the primary cause of most injuries in operating units. The prototype was useful and everyone agreed, in terms of acceptability, drone technology will improve search and rescue operations.

*Conclusion* – The innovative way of utilizing Android and drone technology is a new step towards the improvement of SAR operations in the Philippines.

*Recommendations* – The LiPo battery must be replaced with a higher capacity and the drone operator should undergo a training course and secure a permit from the Civil Aviation Authority of the Philippines (CAAP).

*Social Implication* - Some people are scared of drones due to privacy issues and fears that they could be used to spy against them.

## INTRODUCTION

In search and rescue operations every second counts. To function as efficiently as possible, it is important to be able to obtain a rapid overview of the situation. The type of view that is often only possible from the sky. An Unmanned Aerial Vehicle (UAV) is a type of aircraft that does not require a human pilot onboard. Also known as a drone amongst the general public. Helicopters and airplanes with aerial views are the top choices in aiding search and rescue operations. However, problems occur in most cases where the surveyed area produces blind spots for the camera to capture resulting in slow and occasionally inaccurate searching. Due to their agility, portability, and aerial access advantages, UAVs have already been deployed for Search and Rescue (SAR) operations for several years. By mounting a high-resolution camera on a UAV, it can provide needed aerial imagery for these missions much more cost-effectively than adopting the conventional manned aircraft approach.

Drones locally referred in the country as Remotely Piloted Aircraft Systems (RPAS) (Civil Aviation Authority of the Philippines, n. d.) are regulated under the Civil Aviation Authority of Philippines (CAAP) of the Department of Transportation, crafted a set of rules and regulations that serves as guidance on the registration and operational requirements of drones (Espinola et al., 2019).

*Figure 1.* Comparative analysis

Figure 1 compares the salient features of a traditional aircraft used in search and rescue operations against quadcopter-based Icarus in terms of maneuverability, mobility, maintenance, and surveillance capability.

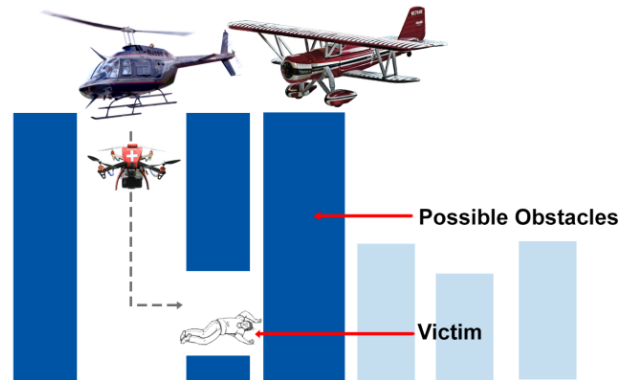

*Figure 2.* Illustrates one of the advantages of using Icarus in civilian applications.

As seen in Figure 2, drones are unmanned, and no additional lives of rescuers in the helicopter are put at risk. With respect to mobility, it can take lesser time to be deployed, fly at low altitudes, and could perform an initial overview of the disaster area. This overview could be better than the one that is done from a helicopter or an aircraft because drones are smaller and can reach places where helicopters cannot.

In addition, drones cost no fuel with the use of Lithium-Ion Polymer (LIPo) batteries. Because drones are smaller than other aircraft performing SAR operations, they also use less energy. This makes them emit fewer greenhouse gasses making it environment friendly.

Time is a crucial factor during any type of search and rescuemission, and to function as efficiently as possible, search and rescue workers obtain a rapid overview of the situation, which is often only possible from the sky. As said by Lean Alfred Santos (2013), for faster and more effective disaster relief and response there is a solution that "flies".

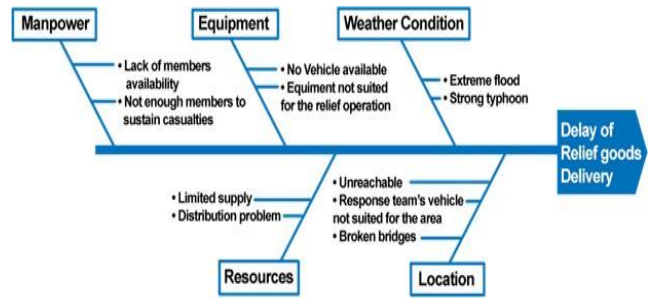

*Figure 3.* The Cause-and-Effect Diagram or Ishikawa Diagram

As described in Figure 3, the common root causes affect delays in the delivery of relief goods during the event of calamities. The diagram is a basic knowledge driven by theoretical and empirical considerations. It is used to map cause and effect relationships, with some more heuristics and some highly quantitative (Kenneth, 2008).

**LITERATURE REVIEW**

*Search, Rescue, and Police Work*

Mayer et al. (2019) claimed that time is frequently the most important consideration in search and rescue situations because lives are in danger. Unmanned aerial vehicle (UAV) deployment by emergency services has already begun to do a faster search over a greater region. In the event of a natural disaster, Mishra et al. (2019) stated that drones can scan the wide affected area and make the search and rescue (SAR) faster to save more human lives. Kaplan and Miller (2019) pointed out that drones are situated in an expanded field of police operations. The alleged distinction between military and civilian, or battlefield and home front.

*Construction Engineering and Engineering*

Tkáč and Mésároš (2019) stated that the most alluring development in building in recent years has been the usage of drones. Drone utilization has increased by approximately 240% in the construction industry, more than in any other commercial sector. Drones have aviation advantages and capabilities that are quite helpful in resolving construction-related issues. According to Nwaogu et al. (2023), the Architecture, Engineering, and Construction (AEC) sector is expected to be the second-

2275

largest market for commercial drones in 2020 due to the sector's relatively quick adoption of the technology.

### *Agriculture, Film, and Television Production*

Upadhyaya et al. (2022) cited that drones are aircraft that can be remotely piloted by a pilot on the ground to carry out a task or that can be automatically flown by loading a flight program that has already been created. Pattanayak and Shukla (2020) conveyed that the increased quality of television news coverage has increased viewers' need for thorough, real-time coverage of breaking news events. In most production regions, the media industry is undergoing a significant transformation. Unmanned aerial vehicles, sometimes known as drones, are emerging as promising new tools for Electronic News Gathering (ENG) and Electronic Field Production (EFP).

### *Medicine and Modern Warfare*

Messar et al. (2018) simulated the delivery of a 4.5 kg load of medical supplies, including tourniquets, bandages, analgesics, and blood products, using an unmanned, rotary-wing drone. The simulated victim was placed in a far-off location. Issacharoff and Pildes (2013) explored the utilization of drones in modern warfare which has become more prevalent. What could have originally been a tactical response, is now a key tactic for attacking the opposition. Braun et al. (2019) suggested that medical drones are on the verge of revolutionizing prehospital medicine enabling advanced healthcare delivery to once-inaccessible patients.

### *News Broadcasting*

Television personality, Daniel Razon (2014), wrote about how UNTV, a news broadcasting company in the Philippines uses drone Technology. According to him, drones are famously used for intelligence operations or aerial assaults. UNTV takes advantage of these unmanned aerial vehicles (UAVs) to intensify its news and rescue missions.

### *Android-Based Application combined with Drone Technology*

Viray (2019) showed other wireless communication methods, such as WiFi, Bluetooth, GSM, or SMS technologies, can be incorporated into a drone to enable wireless data transfer in addition to the usage of remote controls as an electronic drone communication method.

## METHODOLOGY

The purpose of this paper is to develop an Unmanned Aerial Vehicle (UAV) using a quadcopter with the capability of video surveillance, map coordinates, a deployable parachute with a medicine kit or a food pack as a payload, and a collision warning system, remotely controlled, and integrated with an android application to assist in search and rescue operations.

Applied research for the design and development of the functional prototype, quantitative and descriptive statistics to summarize data in an organized manner by describing the relationship between variables in a sample or population (Kaur et al., 2018). Upon completion, the quadcopter underwent an evaluation using a survey instrument with closed-ended questionnaires constructed for data-gathering procedures to limit the respondents' answers to a fixed set of responses (Roopa and Rani, 2012). It validated its acceptability using predefined variables, demonstrated purposively to selected respondents who are involved or part of a rescue unit operating locally within Caloocan City and Quezon City, Philippines. They described some issues and concerns based on their lived experiences in the conduct of search and rescue (SAR) operations, including their demographic profiles.

## *Hardware and Software Components*

Several hardware components were assembled in developing the prototype of Icarus such as F450 frames (450mm), GoolRC A2212/1000kv 13T motors, 30A SimonK brushless Electronic Speed Controller (ESC), Arduino AT Mega 2560 board, Arduino Uno board, FlySky FS-I6 2.4 GHz Radio Transmitter with Receiver, 3s 2200mAh 11.1v Lithium Polymer (LiPo) battery, B3AC 2s-3s charger, HC–SR04 ultrasonic sensors, EMAX 1045 propellers, sensor holders, male to female jumper wires, NRF wireless module, breadboards, light emitting diodes (LEDs), servo motors, 30 cm x 30 cm plastic-made parachute, 4 inches x 4 inches x 4 inches light-weight wooden box, and Pixhawk 4 flight controller. On the software side, Arduino IDE Sketch, Android Studio, and LibrePilot were used in the front-end and coding side. Different analytical tools were also provided as shown below in Figures 4, 5, 6, 7, 8, 9, and 10 to best illustrate how the entire system works collectively through diagrams, flowcharts, and drawn figures.

## *Analytical Tools*

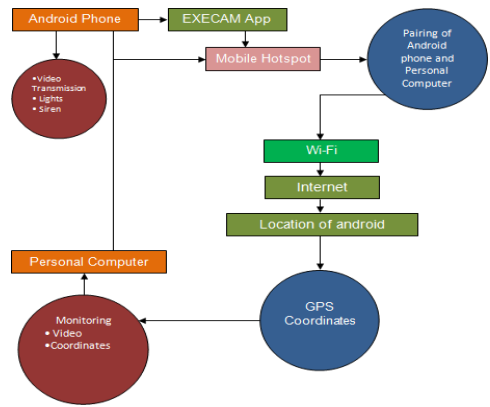

*Figure 4.* Data Flow Diagram of the Android Application

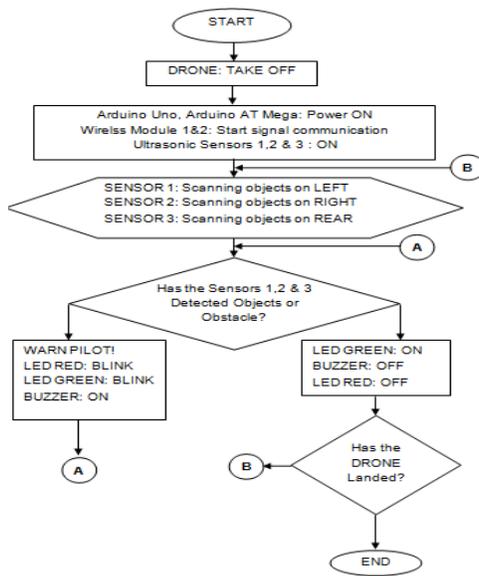

*Figure 5.* System Flowchart of Icarus' collision-detection warning system

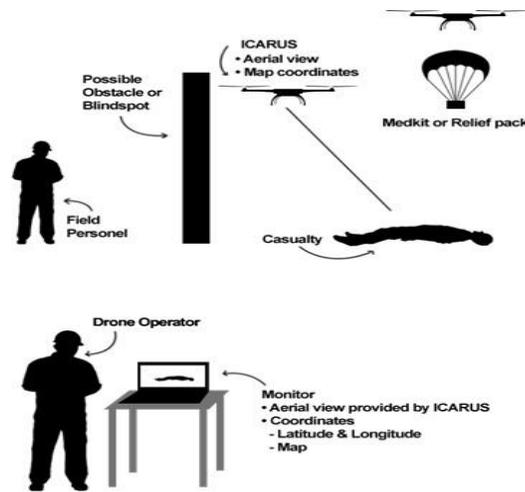

*Figure 6.* Deployment Diagram of Icarus

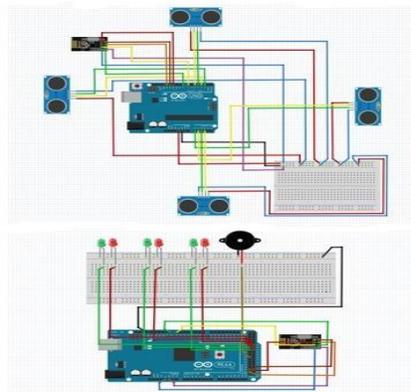

*Figure 7.* Pin diagram of the collision-detection warning system

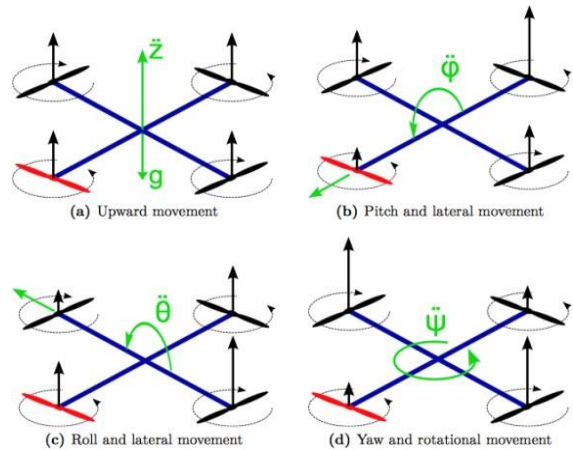

*Figure 8.* How propellers work

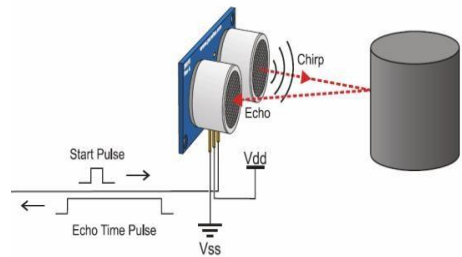

*Figure 9.* How an ultrasonic sensor works

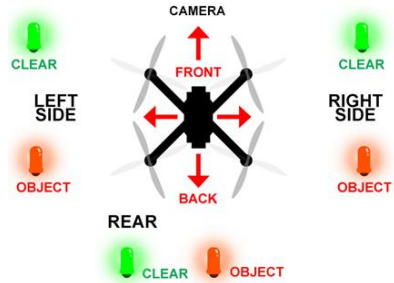

*Figure 10.* The collision-detection warning system with LED indicators

## *Statistical Treatment of Data*

The researchers have employed a simple statistical tool to analyze the data gathered such as variables, frequency distribution (f), percentile (%), and ranked data.

## *The Functional Prototype*

Figure 11 shows the 19 x 12 inches fully functional quadcopter prototype of Icarus. Equipped with 4 brushless motors and propellers in a 450mm frame, mounted with a microcontroller-board connected to a speed and flight controller, wireless modules, an android phone for the navigational system, an ultrasonic sensor, and LED indicators for the collision-detection warning system, powered by a 3s 2200mAh rechargeable 11.1-volts Lithium Polymer (LiPo) battery, remotely controlled by 2.4 GHz receiver and transmitter, fiberglass body is painted with red and white cross color to signify its search and rescue flight mission.

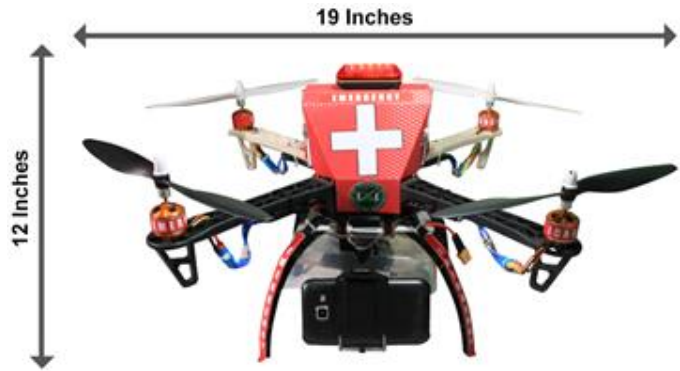

*Figure 11.* Icarus' prototype

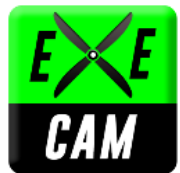

*Figure 12.* ExeCam application icon

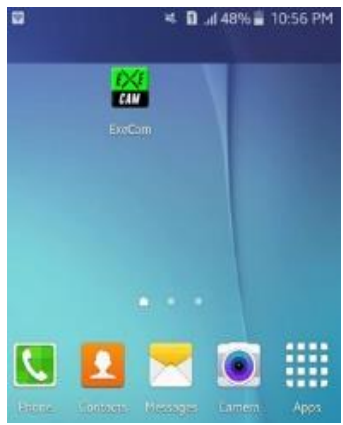

*Figure 13.* ExeCam installed in an Android phone

As seen in Figures 12 and 13, ExeCam, an Android application developed by the researchers has a file size of 3.07 Megabytes and runs on Lollipop 5.1.1 android version.

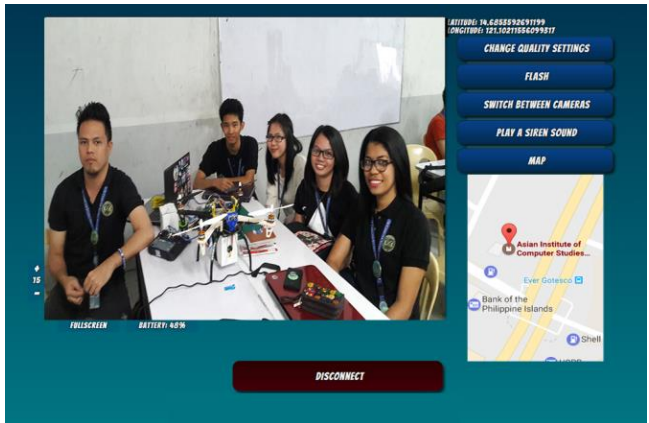

*Figure 14.* Screenshot of ExeCam application

The application uses the Android phone's camera to serve as the medium of video transmission as depicted in Figure 14.

## RESULTS

Tables 1 and 2 summarize the tabulated results of the demographic profiles of the respondents, issues, and concerns distributed accordingly using the statistical tools from highest to lowest that most of the respondents expressed during the actual conduct of the survey in their respective office locations in Caloocan and Quezon City.

Table 1. Distribution of Responses, Frequencies, Percentages, and Ranks
In terms of the Demographic Profile of Respondents

| Variables | Responses | f | % | Rank |
|---|---|---|---|---|
| 1. Respondent's SAR group/unit | Caloocan City Disaster Risk and Monitoring Office | 10 | 33.3 | 1 |
| | New Era Bureau of Fire Protection | 10 | 33.3 | 1 |
| | Quezon City Police District Station 6 | 10 | 33.3 | 1 |
| 2. Gender | Male | 27 | 90 | 1 |
| | Female | 3 | 10 | 2 |
| 3. Civil Status | Single | 16 | 53 | 1 |
| | Married | 14 | 47 | 2 |
| | Widow | 0 | 0 | |
| 4. Age Group | 31 years old - Above | 11 | 37 | 1 |
| | 21 – 25 years old | 9 | 30 | 2 |
| | 26 – 30 years old | 7 | 23 | 3 |
| | 18 – 20 years old | 3 | 10 | 4 |

Table 2. Distribution of Responses, Frequencies, Percentages, and Ranks
In terms of Issues and Concerns

| Variables | Responses | f | % | Rank |
|---|---|---|---|---|
| 1. Types of SAR Operation | Ground SAR | 13 | 43 | 1 |
| | Firefight SAR | 10 | 33 | 2 |
| | Urban SAR | 5 | 17 | 3 |
| | Combat SAR | 2 | 7 | 4 |
| | Ground SAR | 13 | 43 | 1 |
| 2. Common Causes of Injuries during SAR Operations | Human Errors | 24 | 80 | 1 |
| | Location Difficulty | 5 | 17 | 2 |
| | Harsh Weather | 1 | 3 | 3 |
| 3. Describing the Prototype | Useful | 15 | 47 | 1 |
| | Reliable | 6 | 20 | 2 |
| | Unique | 6 | 20 | 2 |
| | High quality | 2 | 10 | 3 |
| | Impractical | 1 | 3 | 4 |
| 4. Quality of the Prototype | Ground SAR | 13 | 43 | 1 |
| | Firefighter | 10 | 33 | 2 |
| | Urban SAR | 5 | 17 | 3 |
| | Combat SAR | 2 | 7 | 4 |
| 5. The Integration of Android Technology | Useful | 23 | 77 | 1 |
| | Reliable | 5 | 17 | 2 |
| | Unique | 2 | 6 | 3 |
| 6. Number of Years in SAR Operations | 1-2 years | 15 | 50 | 1 |
| | 3-4 years | 5 | 17 | 2 |
| | 9-10 years | 3 | 10 | 3 |
| | 5-6 years | 3 | 10 | 3 |
| | 13-14 years | 2 | 7 | 4 |
| | 15-20 years | 1 | 3 | 5 |
| | 7-8 years | 1 | 3 | 5 |
| 7. Success Rate of SAR Operations | 90 – 100 % | 15 | 50 | 1 |
| | 70 – 89 % | 9 | 30 | 2 |
| | 50 – 69 % | 3 | 10 | 3 |
| | 30 – 49 % | 3 | 10 | 3 |
| 8. Acceptability of Drone Technology in SAR Operation | Agree | 30 | 100 | **1** |
| | Disagree | 0 | 0 | |

## DISCUSSION

Based on Tables 1 and 2, arranged from highest to lowest concerning frequencies and percentages, while in ranks, number 1 being the highest, the researchers analyzed and interpreted the results in terms of the demographic profile of the respondents and issues and concerns.

Each SAR group/unit was composed equally of 33.3% or 10 members each from "The Caloocan City Disaster Risk and Monitoring Office (CCDRMO)", "New Era Bureau of Fire Protection (New Era

BFP)", and "Quezon City Police District 6 (QCPD 6)". There were 27% "Male" and 3% "Female".

Without being gender-biased, men overpower women this is due to the fact of the nature of the job where mental and physical strength are the primary requirements to carry out the task. 53% were "Single" and 47% with "Married" civil status.

The slight difference between the number of singles and married are not considered a factor in being flexible and available during the conduct of SAR operations.

37% belonged to the "31-years old – above" age group, followed by 30% "25-26 years old", 23% "26-30 years' old, and 10% "18-20 years old".

"Ground SAR" is selected by 43% of respondents. This rescue effort focuses on searching for a lost or distressed person in a given location, followed by "Firefight SAR" with 33% concerned with saving people trapped inside a burning structure, "Urban SAR" with 17%, involves rescue operations where victims are trapped in collapsed structures due to natural and man-made disasters, and "Combat SAR" with 7%, carries out search and rescue missions in war conflict territories or zones.

80% of the common causes of injuries are listed as "Human Errors", including impaired judgment, lack of training/experience, mental block, distraction, arrogance, and overconfidence, by 17% as "Location Difficulty", this is due to unfamiliarity with the terrain, and 13% as "Harsh Weather Conditions", circumstances in which climate are extremes and unexpected.

Icarus is described as "Useful" by 47% of respondents, "Reliable" by 20%, "Unique" by 20%, and "High Quality" by 10%, while just 3% of respondents find it "Impractical."

To sum up, 97% of respondents affirmatively described the potential of Icarus against the 3% negative responses. Overwhelming 97% of respondents rated the project as "High Quality," whereas 3% rated it as "Poor Quality."

The 3% poor quality rate is manageable and can be improved further. 77% of respondents found Android technology to be "Useful," 17% "Reliable," and 6% "Unique." The 100% positivity rate is quite impressive. 50% of the respondents have "1-2 years," followed by 17% with "3-4 years," 10% between "5-6 years" and "7-8 years," 7% with "13-14 years," and both at 3% with "15-20 years" and "7-8 years."

It goes to show that SAR operators with 1-2 years of experience are equal in number to those with 3-20 years of experience. 50% of operations had a success rate of "90% to 100%". 30% between "70-89 percent," 10% between "30-49 percent," and 50% between "50-69 percent," respectively. The success is relatively high from 50% to 100%.

The overall acceptability rate on the use of drone technology in search and rescue operations was 100% unanimously "Agreed" by all respondents.

## CONCLUSIONS AND RECOMMENDATIONS

The data provided by search and rescue operators enabled researchers to recognize the importance and significance of using Android and drone technology to support search and rescue operations. In terms of the overall survey results, most of the respondents asserted a positive response in describing the answers to all the issues and concerns.

Therefore, the researchers concluded that the innovative way of utilizing android and drone technology is a new step toward the improvement of search and rescue operations in the Philippines.

The LiPo battery must be replaced with a higher capacity to extend longer flight time, drone operator should undergo a proper drone training course, and secure a permit from the Civil Aviation Authority of the Philippines (CAAP) to ensure public safety.

## SOCIAL IMPLICATIONS

Some people are scared of drones. They said that their right to privacy might be in danger, and there are fears that drones could be used to spy on them. Since drones were deployed as a killing machine in the Russian-Ukrainian conflict that began on February 24, 2022, any SAR operations could be affected by the public's negative perception of drones.

## ACKNOWLEDGEMENT

Always grateful to the men and women of Caloocan City Disaster Risk and Monitoring Office, New Era Bureau of Fire Protection, Quezon City Police District Station 6, and most especially to Ma. Lyn M. Feliciano, April Joy E. Capilitan, and Samira B. Gumbahale for extending much-needed assistance and providing moral support. Lastly, the Asian Institute of Computer Studies and the Philippine State College of Aeronautics for making the collaboration possible.

## DECLARATIONS

### *Conflict of Interest*

All authors declared that they have no conflicts of interest.

### *Informed Consent*

All participants were appropriately informed and voluntarily agreed to the terms with full consent before taking part in the conduct of the experiment/survey.

*Ethics Approval*

The AICS Research Ethics Committee duly approved this study after it conformed to the local and internationally accepted ethical guidelines.

## Author's Biography

**Dr. Manuel Luis C. Delos Santos** holds a bachelor's degree in the Bachelor of Science in Computer Science from the Divine Word College of Vigan in Ilocos Sur, a Master of Science in Computer Science, and a Doctor in Information Technology taken both from AMA University, Quezon City. He is the current Dean of the Bachelor of Science in Computer Science degree program at the Asian Institute of Computer Studies (AICS) located in Quezon City, Philippines.

**Dr. Jerum B Dasalla** obtains his Bachelor of Science Major in Computer Science from the Emilio Aguinaldo College, Manila, a Master of Education in Aeronautical Management at the Philippine State College of Aeronautics, and a master's in information technology and a Doctor of Information Technology taken both from AMA University, Quezon City. He is the current On-the-Job Training Coordinator of the Bachelor of Science in Aviation Information Technology and Bachelor of Science in Aviation Information System degree programs at the Philippine State College of Aeronautics (PhilSCA), Pasay City.

**Jomar C. Feliciano** is a Computer Science graduate with a strong inclination towards research and robotics, combining his knowledge and expertise. He currently leads as the President of Ocean Hive PH (oceanhiveph.com), a thriving start-up specializing in mobile and web development in Cebu City, Philippines.

**Dustin Red B. Cabatay** is a talented Software Engineer who finished a bachelor's degree in computer science and has a deep passion for innovation and problem-solving. His experience in cutting-edge software solutions has contributed to numerous successful projects, showcasing his expertise in programming languages and software development methodologies. His dedication and drive make him a valuable asset in any technological endeavor.